\documentclass[amsmath,amssymb,bibnotes]{revtex4}

\usepackage{color,epsfig,rotating,graphicx,subfigure}

\usepackage{amsmath,amssymb,amscd}

\usepackage[latin1]{inputenc}
\usepackage[english]{babel}

\usepackage{dsfont}
\renewcommand{\Im}{{\rm Im}}

\newcommand{\re}{{\rm e}}
\newcommand{\rd}{{\rm d}}

\newcommand{\kb}{k_{\rm B}}

\newcommand{\Tr}{{\rm Tr}}

\begin{document}
\title{Scalable radiative thermal logic gates based on nanoparticle networks}

\date{\today}

\author{Christoph Kathmann$^1$, Marta Reina$^2$, Riccardo Messina$^2$, Philippe Ben-Abdallah$^2$, Svend-Age Biehs$^{1,*}$}
\affiliation{$^1$ Institut f\"{u}r Physik, Carl von Ossietzky Universit\"{a}t, D-26111 Oldenburg, Germany}
\affiliation{$^2$ Laboratoire Charles Fabry,UMR 8501, Institut d'Optique, CNRS, Universit\'{e} Paris-Sud 11,
2, Avenue Augustin Fresnel, 91127 Palaiseau Cedex, France}
\affiliation{$^*$Corresponding author: s.age.biehs@uni-oldenburg.de}

\begin{abstract}  
We discuss the design of the thermal analog of logic gates in systems made of a collection of nanoparticles. We demonstrate the possibility to perform  NOT, OR, NOR, AND and NAND logical operations at submicrometric scale by controlling the near-field radiative heat exchanges between their components. We also address the important point of the role played by the inherent non-additivity of radiative heat transfer in the combination of logic gates. These results pave the way to the development of compact thermal circuits for information processing and thermal management.
\end{abstract}

\maketitle

In electronics a logic gate is a circuit which implements a Boolean operation from input digits based on an electric signal with distinct discrete levels. Non-electronic devices have been proposed during the last decades to perform such Boolean treatment  without electric power source. Hence, purely optical~\cite{Tsai},  mechanical~\cite{Singh} and even biological devices~\cite{Endy,Moe} have been developped to process information without electric current. In 2007 the idea of basic thermal logic gates~\cite{BaowenLi1} has been introduced, in order to make logical operations using temperature differences and heat flux carried by acoustic phonons in non-linear solid circuits as an alternative to electric circuits. 

Since the beginning of {2000's}, the possibility to control the radiative heat exchanges at the nanoscale between two or several objects mechanically~\cite{GhanekarEtAl2018,BiehsEtAl2011,ElzoukaNdao2017}, electrically and chemically~\cite{HuangEtAl2014,MessinaEtAl2017,IlicEtAl2018,ShiEtAl,YiEtAl2019}, or with external magnetic fields \cite{MoncadaVilla2015,Zhu2016,Latella2017,Cuevas,BenAbdallah2016,OttEtAl2019a, WuEtAl2019} has been demonstrated, opening so promising prospects in this domain as well as in the field of thermal management.

As a result, non-linear two-body devices have been first proposed to rectify heat flux both in near-field and in far-field regimes~\cite{Zheng2011,Iizuka2012,PBASAB2013,Yangetal2013,ItoEtAl,ItoEtAl2017,FiorinoEtAl2018,FanRectification2010,BasuEtAl2011,WangEtAl2013,Nefzaoui2014,OrdonezEtAl2017,OttEtAl2019b}, allowing thus the fabrication of true radiative thermal diodes. More recently, three-terminal systems have unveiled the possibility to store and to amplify the thermal energy carried by thermal photons opening the door to the realization of radiative memories and transistors~\cite{PBASAB2013,Yangetal2013,PBASAB2014,OrdonezEtAl2016,Kubytskyi,Dyakov,Rodriguez,Ivan2019}. This paves the way to thermal circuits to control heat flow at nanoscale in a similar way as in modern electronics~\cite{PBASAB2016}.

These recent results, promising in the direction of information treatment with thermal photons and thermal management in general, suffer of two main limitation. The first is that the planar geometry considered so far does not leave much room for both down-sizing of the device and geometrical arrangement in order to combine different ports and realize more elaborate thermal circuits. The second limitation is that only a NOT, OR and AND gates have been demonstrated so far. In order to overcome the first drawback, we address here the problem of the design of thermal logic gates based on nanoparticles. These systems, which are nowadays experimentally feasible, remove the constraint of a 1D arrangement, paving the way to an easier combination of logic ports. Moreover, the dipolar approximation allows for a relatively easy theoretical treatment of heat exchange in these systems. For this configuration, we extend our previous work on basic logic ports and we design NOT, OR, NOR, AND and NAND logic ports. We also discuss the important problem of how the inherent non-additivity of radiative heat transfer affects the combination of several ports together and address the scalability of the designed logic ports.

{\section{Methods}}

\subsection{Theoretical framework}

To explore theoretically the possibility to realize different logic gates and their combination we restrict ourselves to many-body systems consisting of $N$ spherical nanoparticles in vacuum, for convenience. These nanoparticles will serve as the input/output terminals of the logic gates and are treated within the dipole model which can be safely used when the particle radii are much smaller than the interparticle distances~\cite{Becerril2019}. In this case, the power received by a particle $i$ at a temperature $T_i$ through heat radiation from particle $j$ with temperature $T_j$ in the presence of all other $N-2$ particles can be expressed 
by~\cite{MessinaEtAl2013,DongEtAl2017,EkerothEtAl2017}
\begin{equation}
  P_{j \rightarrow i} = \int_0^\infty \!\! \frac{\rd \omega}{2\pi}\, \frac{\Im(\alpha_i)\Im(\alpha_j)}{|\alpha_i|^2} \Theta_{ji} \Tr\biggl( \mathds{T}^{-1}_{ij} {\mathds{T}^{-1}_{ji}}^\dagger \biggr),
\label{Eq:Pji}
\end{equation}
with
\begin{equation}
  \Theta_{ji} = \frac{\hbar \omega}{\re^{\hbar \omega/\kb T_j} - 1} - \frac{\hbar \omega}{\re^{\hbar \omega/\kb T_i} - 1},
\end{equation}
introducing the reduced Planck constant $\hbar$ and the Boltzmann constant $\kb$. The transferred power depends on the product of the imaginary parts of the two particles. For a given spherical particle $i$ we take the Clausius-Mossotti expression of the polarizability
\begin{equation}
  \alpha_i = 4 \pi a^3_i \frac{\epsilon_i - 1}{\epsilon_i + 2}
\end{equation}
where $a_i$ is the radius and $\epsilon_i$ is the permittivity of particle $i$. Furthermore the exchanged power depends on the matrix $\mathds{T}$ which is a $3N\times3N$ block matrix with components defined by~\cite{MessinaEtAl2013}
\begin{equation}
  \mathds{T}_{ij} = \delta_{ij} \mathds{1} - (\delta_{ij} - 1) \frac{\omega^2}{c^2} \alpha_i \mathds{G}^{(0)}(\mathbf{r}_i,\mathbf{r}_j).
\end{equation}
Here, $\mathds{G}^{(0)}(\mathbf{r}_i,\mathbf{r}_j)$ is the well-known Green's dyadic in vacuum~\cite{Hecht} evaluated at the 
positions $\mathbf{r}_{i/j}$ of particle $i/j$ and $c$ is the vacuum light velocity. Effects like the coupling to surface modes of a nearby substrate as studied in Refs.~\cite{Saaskilathi2014,Asheichyk2017,DongEtAl2018,MessinaEtAl2018} are therefore neglected at this stage. Furthermore, we neglect the heat flux between the particles and the surrounding environment, which itself provides a heat flux channel, since this gives a negligible contribution when the particles are interacting in the near field~\cite{MessinaEtAl2013}. In agreement with this simplification we also neglect the radiation correction~\cite{AlbadalejoEtAl2010}. More important in the following is that through the block matrix $\mathds{T}$ we take into acount all many-body interactions. Note, that the polarizability $\alpha$ is proportional to the particle radius $a^3$ and the Green's function scales in the near-field regime like $1/d^3$ where $d = |\mathbf{r}_i - \mathbf{r}_j|$ is the distance between two particles. Therefore the second term  in the matrix $\mathds{T}$ scales like $(a/d)^3$ in the near-field regime. Hence, when scaling the radii of all particles and the distances by the same factor, the results remain unchanged. Therefore the whole systems considered in the following can be down- and upscaled without changing the functionalities by multiplying the distances and radii with one scaling factor.  {The limit for the downscaling is, of course, the validity of the macroscopic approach which can be applied to length scales much larger than the atomic dimensions or the lattice constant of the material where the dielectric constant cannot be defined anymore. Practically speaking this occurs for particles of sub-nanometer radius where quantum effects must be taken into account. The limit for the upscaling is the validity of the dipole approximation and near-field regime. Both are typically valid for radii and distances smaller than 1/10 of the thermal wavelength. For logic gates operating around the ambient temperature this corresponds to sizes larger than a few micrometers. However the higher-order modes (i.e. multipolar contributions) can easily be taken into account to deal with bigger particles. Also the well known proximity approximation could be used to deal with big particles.}

Finally, the full power received or emitted by the particle $i$ is given by 
\begin{equation}
  P_{i}(T_1,\ldots,T_N) = \sum_{j\neq i} P_{j \rightarrow i}.
\end{equation}
Note that the dipole model offers the possiblity to model also more complex geometries for the input and output terminals in a discrete-dipole approximation~\cite{MessinaEtAl2013,DongEtAl2017,EkerothEtAl2017,PBAEtAl2011,Edalatpour2014,Edalatpour2015} which is a huge advantage regarding the constraints imposed by layered media as considered for the construction of transistors and logic gates previously~\cite{PBASAB2014,OrdonezEtAl2016,PBASAB2016}. Furthermore, the discrete dipole approximation also offers the possibility to design complex and nearly arbitrary circuits for thermal radiation.

\subsection{Optical properties}

\begin{figure}
	\centering
	\includegraphics[width=0.4\textwidth]{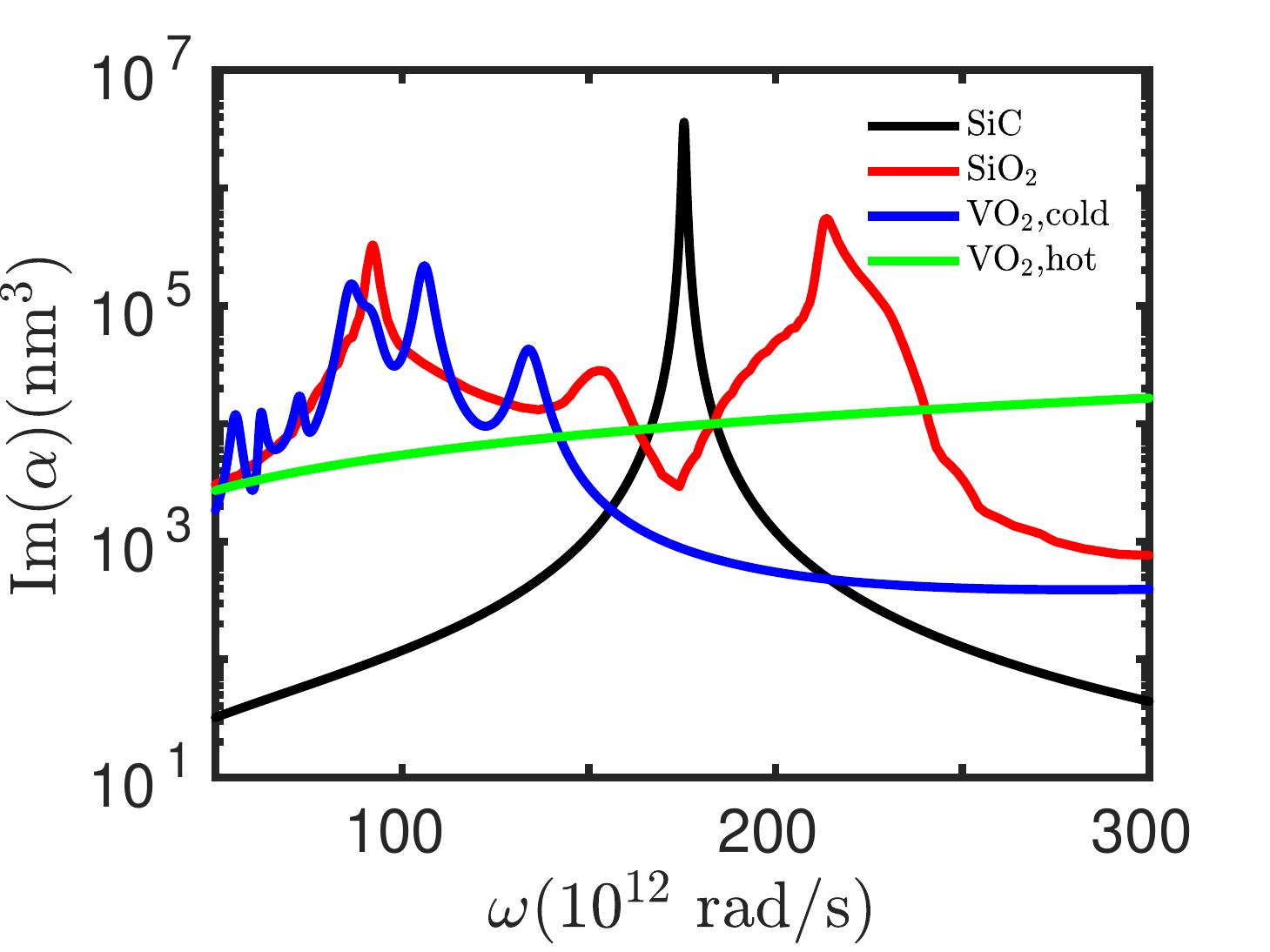}
	\caption{{Particle polarizabilities.} Plot of $\Im(\alpha)$ for spherical SiC and VO$_2$ nanoparticles with radius $a = 25\,{\rm nm}$.}
	\label{Fig:ImagAlpha}
\end{figure}

In our simulations we will use the optical properties of VO$_2$, SiO$_2$ and SiC from literature~\cite{Palik,QazilbashEtAl2007}. These optical properties determine the absorptivity of the particles which is proportional to $\Im(\alpha_{i})$ and the coupling strengths which are proportional  to $\Im(\alpha_i) \Im(\alpha_j)$. In Fig.~\ref{Fig:ImagAlpha} we show $\Im(\alpha_i)$ of SiC, SiO$_2$ and VO$_2$ nanoparticles. The absorptivity of the SiC particle has a sharp resonance at the position of the localized surface phonon polariton at $\omega = 1.756 \times 10^{14}{\rm rad}/{\rm s}$. Similarly, SiO$_2$ shows peaks at the phonon polariton resonances around $\omega = 1 \times 10^{14}{\rm rad}/{\rm s}$, $\omega = 1.5 \times 10^{14}{\rm rad}/{\rm s}$, $\omega = 2.1 \times 10^{14}{\rm rad}/{\rm s}$.  Now, it is important to note that VO$_2$ has a sharp phase transition at a critical temperature $T_{\rm c} = 340\,{\rm K}$ from a dielectric ($T < T_{\rm c}$) to a metallic state ($T > T_{\rm c}$). In its dielectric state it also supports localized surface mode resonances clearly seen in the spectrum of $\Im(\alpha)$. In the metallic state the absorptivity shows no resonance feature, i.e.\ there are no localized surface mode resonances in the shown frequency region. It is interesting that for frequencies larger than $1.5\times10^{14}\,{\rm rad}/{\rm s}$ the absorptivity of VO$_2$ in its metallic state is larger than that of VO$_2$ in its dielectric state. Furthermore, in the frequency region below $\omega = 1.5 \times 10^{14}{\rm rad}/{\rm s}$ the phonon-polariton resonances of VO$_2$ in its dielectric state nicely overlap with the ones of SiO$_2$.

{\section{Results and Discussion}}

\subsection{NOT and UNIT gate}

\begin{figure}
	\centering
	\includegraphics[width=0.4\textwidth]{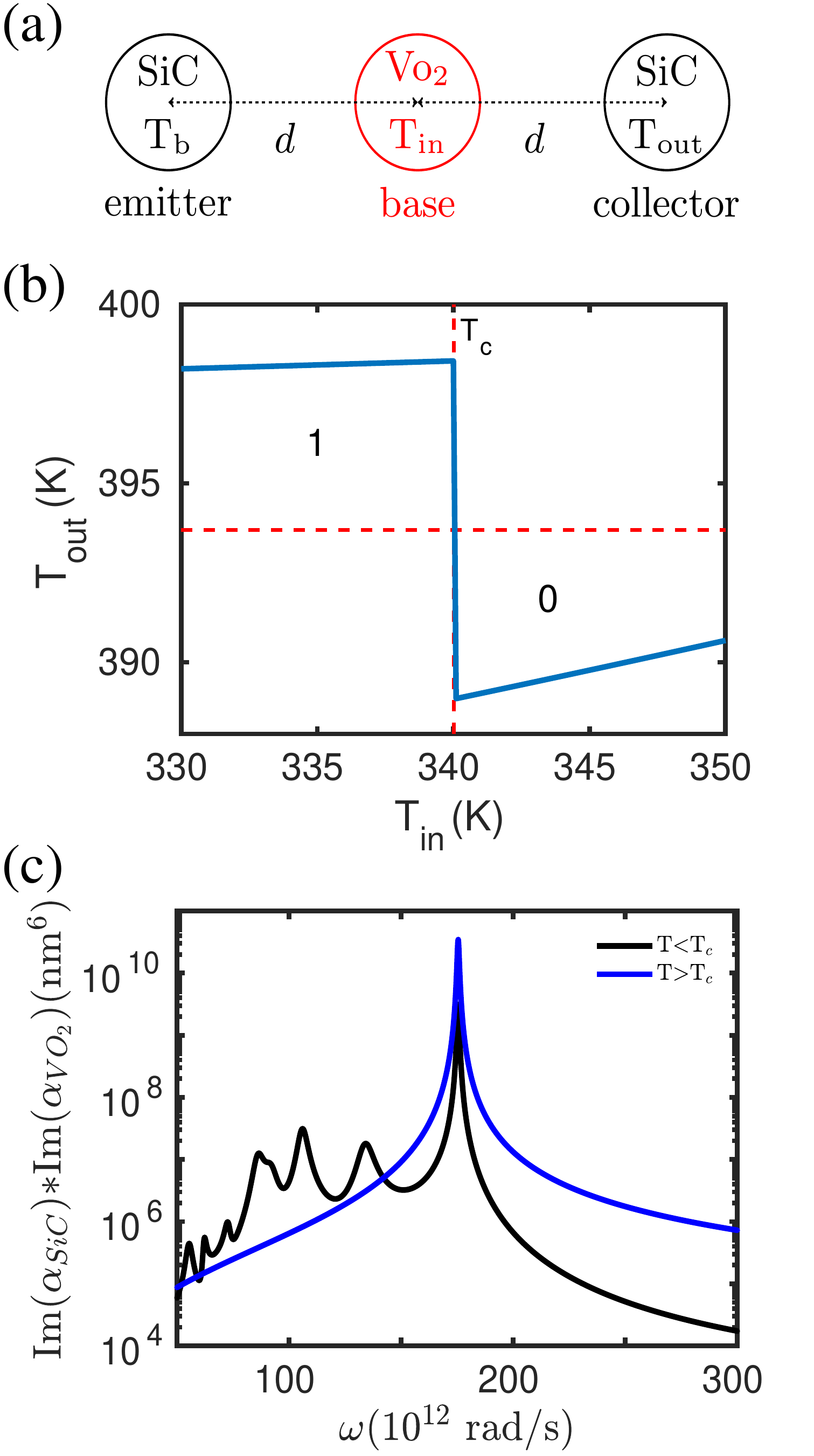}
	\caption{ {NOT gate.} (a) Sketch of the NOT gate configuration. The radius of the nanoparticles is $a = 25\,{\rm nm}$ and the interparticle distance is $d = 100\,{\rm nm}$. (b) Output of the NOT gate in (a) when $T_{b} = 400\,{\rm K}$. The ``0" (``1") of the input is defined by $T_{\rm in} < T_{\rm c}$ ($T_{\rm in} > T_{\rm c}$) . The ``0" (``1") of the output is defined by $T_{\rm out} < 394\,{\rm K}$ ($T_{\rm out} > 394\,{\rm K}$). (c) $\Im(\alpha_{\rm SiC})\Im(\alpha_{{\rm VO}_2})$ for VO$_2$ in the dielectric (metallic) phase for which $T < T_{\rm c}$ ($T > T_{\rm c}$).}
  \label{Fig:NOTgate}
\end{figure}

We will now use Eq.~(\ref{Eq:Pji}) to simulate the functionalities of logic gates with input/output terminals represented by a nanoparticle. We start with a realization of a NOT gate. To this end, we consider a three-body configuration as depicted in Fig.~\ref{Fig:NOTgate}(a) where the collector  and the emitter are given by SiC nanoparticles and the base is a VO$_2$ particle. To operate this NOT gate, we fix the temperature of the emitter at $T_b = 400\,{\rm K}$ and take the base particle as the input terminal and the collector particle as the output terminal, i.e.\ we control the input temperature $T_{\rm in}$ and let the output particle relax into its nonequilibrium steady state. The corresponding output temperature $T_{\rm out}$ is determined by the condition that the full power received or emitted by the collector particle is zero, which can be evaluated numerically, e.g. using Newton's method, to determine $T_{\rm out}$ such that $P_{\rm collector} (T_{\rm b}, T_{\rm in}, T_{\rm out}) = 0\,{\rm W}$. The results for $T_{\rm out}$ as a function of $T_{\rm in}$ are shown in Fig.~\ref{Fig:NOTgate}(b). It can be seen that at $T_{\rm in} = T_{\rm c} = 340\,{\rm K}$ there is a jump of the output temperature from $398\,{\rm K}$ to $388\,{\rm K}$. Hence, if we define input temperatures with $T_{\rm in} < 340\,{\rm K}$ as ``0" and $T_{\rm in} > 340\,{\rm K}$ as ``1" and correspondingly $T_{\rm out}< 394\,{\rm K}$ as ``0" and $T_{\rm out} > 394\,{\rm K}$ as ``1" then we have a realization of a NOT gate. 

To understand the physics behind the NOT gate operation, we show in Fig.~\ref{Fig:NOTgate}(c) the quantity $\Im(\alpha_{{\rm SiC}_2}) \Im(\alpha_{{\rm VO}_2})$ for the case where the VO$_2$ particle is in its dielectric state ($T < T_{\rm c}$) and for the metallic state ($T > T_{\rm c}$). It can be seen that because $\Im(\alpha_{{\rm VO}_2})$ is for $T > T_{\rm c}$  larger than for $T < T_{\rm c}$ at the resonance of the SiC particle in Fig.~\ref{Fig:ImagAlpha} the coupling between the VO$_2$ particle and the SiC particle is stronger for $T > T_{\rm c}$ than for $T < T_{\rm c}$ as can be seen in Fig.~\ref{Fig:NOTgate}(c). Therefore, when the VO$_2$ particle is in its dielectric phase, the cooling of the output particle by the VO$_2$ input particle is small. On the other hand, when the  VO$_2$ particle is in its metallic phase the input VO$_2$ particle has a significant cooling effect on the output particle. This explains why the output particle is hotter for $T_{\rm in} < T_{\rm c}$ than for  $T_{\rm in} > T_{\rm c}$. The NOT ability is therefore provided by the phase-change of the VO$_2$ input particle and the corresponding change of the coupling to the output particle. 

\begin{figure}
	\centering
	\includegraphics[width=0.4\textwidth]{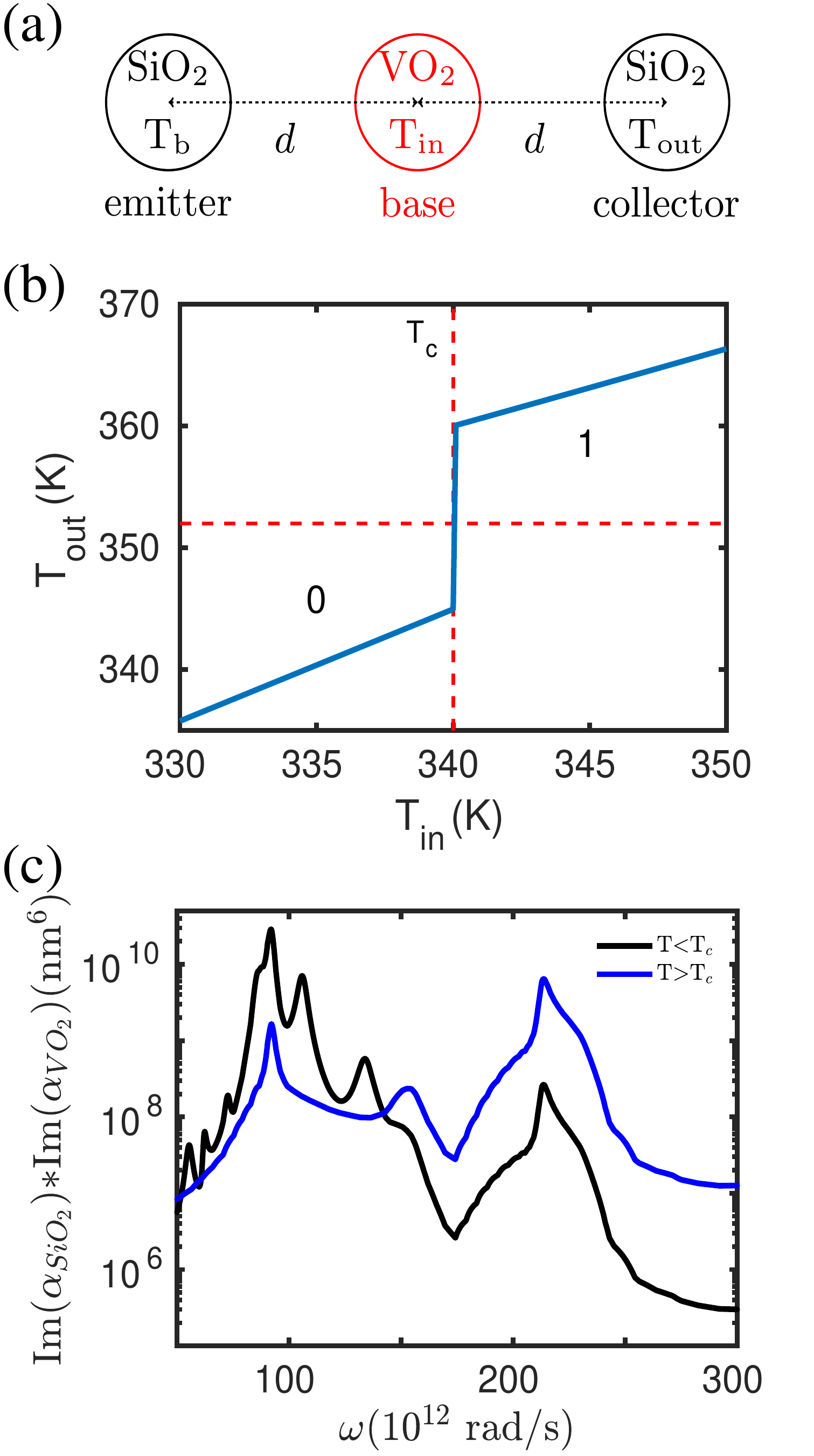}
	\caption{{UNIT gate.} (a) Sketch of the UNIT gate configuration. The radii of the nanoparticles is $a = 25\,{\rm nm}$ and the interparticle distance is $d = 100\,{\rm nm}$. (b) Output of the UNIT gate in (a) when $T_{b} = 400\,{\rm K}$. The ``0" (``1") of the input is defined by $T_{\rm in} < T_{\rm c}$ ($T_{\rm in} > T_{\rm c}$) . The ``0" (``1") of the output is defined by $T_{\rm out} < 352\,{\rm K}$ ($T_{\rm out} > 352\,{\rm K}$). (c) $\Im(\alpha_{{\rm SiO}_2})\Im(\alpha_{{\rm VO}_2})$ for VO$_2$ in the dielectric (metallic) phase for which $T < T_{\rm c}$ ($T > T_{\rm c}$ ).}
  \label{Fig:NOTgateGlass}
\end{figure}

By replacing now the SiC nanoparticles by SiO$_2$ nanoparticles as shown in Fig.~\ref{Fig:NOTgateGlass}(a), we observe in Fig.~\ref{Fig:NOTgateGlass}(b) that this configuration functions as a UNIT gate. Again this behaviour can be understood by the coupling strength $\Im(\alpha_{{\rm SiO}_2})\Im(\alpha_{{\rm VO}_2})$ visualized in Fig.~\ref{Fig:NOTgateGlass}(c) where it can be seen that the peaks around $\omega = 1\times10^{14}\,{\rm rad/s}$ are much larger for $T_{\rm in} < T_c$ than for $T_{\rm in} > T_c$. That means, in contrast to the SiC particles, that the SiO$_2$ particles are well coupled with the VO$_2$ particles when $T < T_{\rm c}$ due to the overlap of the localized resonant modes in both particles. On the other hand, for $T > T_{\rm c}$ the particles are not well coupled so that for $T < T_{\rm c}$ the VO$_2$ input particle efficiently cools down the output SiO$_2$ particle, whereas for $T > T_{\rm c}$ the cooling of the VO$_2$ input particle becomes less efficient resulting in the UNIT gate functionality. 

It is now interesting to observe that, by operating the NOT gate slightly differently, we can invert its functionalities. More specifically, the output of the NOT gate configuration in Fig.~\ref{Fig:NOTgate}(a) is completely changed when the emitter particle temperature $T_b$ is changed from $400\,{\rm K}$ to  $300\,{\rm K}$. The output temperature of the NOT gate with $T_b = 300\,{\rm K}$ are shown in Fig.~\ref{Fig:Fig4}(a). We clearly see that the functionality is inverted and the NOT gate operates as a UNIT element with respect to the logical operation when defining $T_{\rm out}< 306\,{\rm K}$ as ``0" and $T_{\rm out} > 306\,{\rm K}$ as ``1". This new functionality can again be understood from the coupling between the VO$_2$ and SiC particles. The output particle is not well coupled to the input particle for $T_{\rm in} < T_{\rm c}$ so that the emitter particle mainly cools the output particle whereas the heating by the presence of the input particle is small. For $T_{\rm in} > T_{\rm c}$ the coupling between the input particle and the output particle is stronger and the heating of the output particle by the input particle is enhanced so that $T_{\rm out}$ makes an upward jump for $T_{\rm in} > T_{\rm c}$. We conclude that the transition between the two opposite ports NOT and UNIT can be realized both optically (by using a different nanoparticle) and by acting on the emitter particle temperature $T_b$, which can thus be seen as a control parameter. 

\subsection{OR gate}

\begin{figure}
	\centering
	\includegraphics[width=0.45\textwidth]{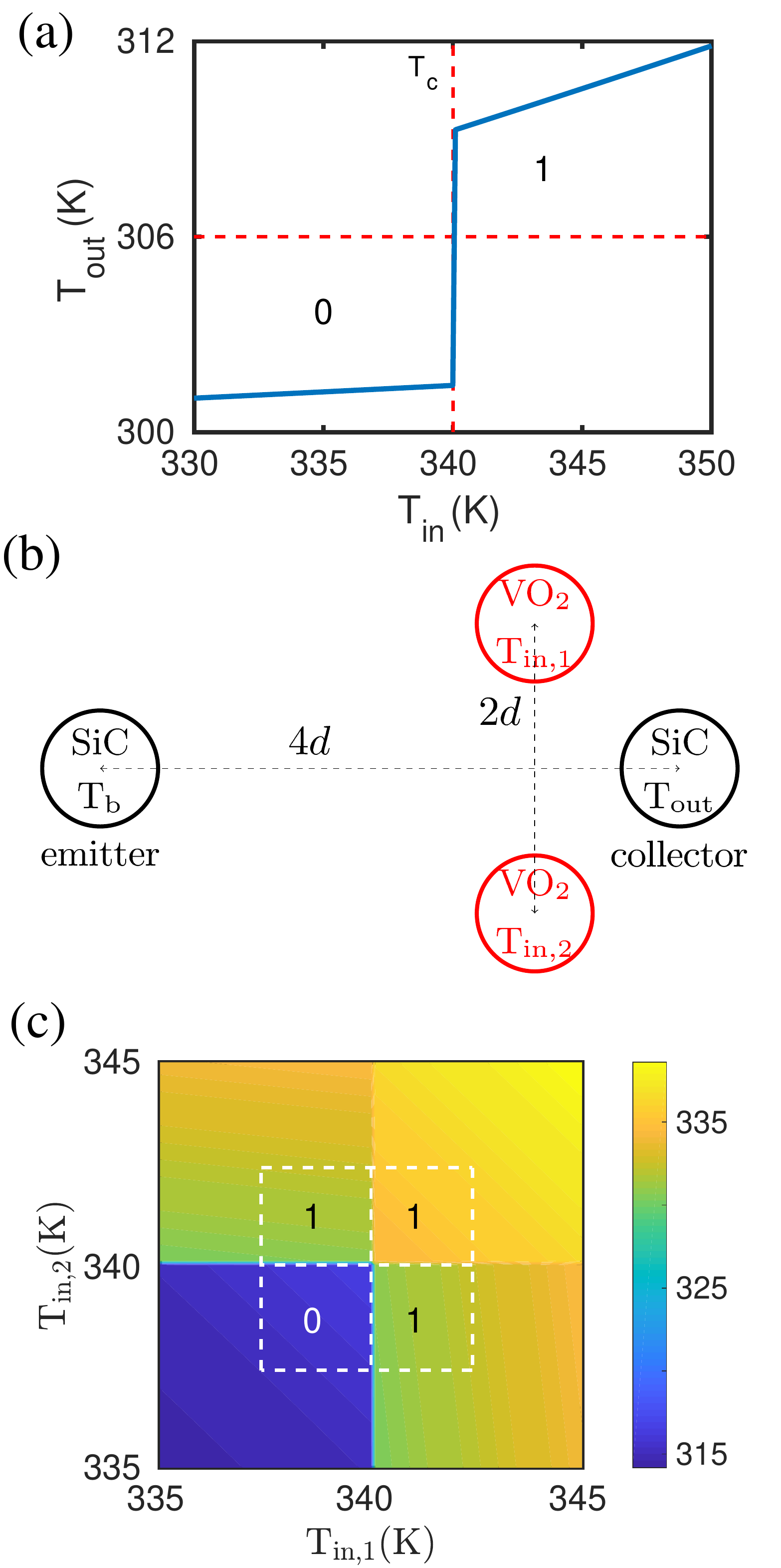}
	\caption{{UNIT and OR gate.} (a) The output temperature of the configuration in Fig.~\ref{Fig:NOTgate}(a) with $T_b = 300\,{\rm K}$. The functionality is that of a UNIT gate for logical operations. (b) Sketch of the OR gate configuration. The radii of the nanoparticles is $a = 25\,{\rm nm}$ and the interparticle distance between the two input particles is $2d = 200\,{\rm nm}$ and between the emitter and collector particle the distance is $4d = 400\,{\rm nm}$ and $T_b = 400\,{\rm K}$. (c) The output temperature of the configuration in (a) with $T_b = 300\,{\rm K}$. The functionality is that of an OR gate.}
  \label{Fig:Fig4}
\end{figure}

We now move to the realization of another basic port, namely the OR logic gate. To this aim we start from the second realization of the UNIT gate, namely the one given in Fig.~\ref{Fig:NOTgate}(a) with $T_b=300\,$K, and replace the single VO$_2$ central particle with two VO$_2$ particles, obtaining the geometrical configuration described in Fig.~\ref{Fig:Fig4}(b). This is equivalent to having two UNIT gates in parallel. In this configuration, the temperature of the output particle is shown in Fig.~\ref{Fig:Fig4}(c). Since the functionalities of the UNIT gate works for the two input particles separately, the output is clearly that of an OR gate if we define the $T_{\rm out} > 325\,{\rm K}$ as ``1" and  $T_{\rm out} < 325\,{\rm K}$ as ``0".

\subsection{NOR gate: combination of two ports vs independent realization}

\begin{figure}
	\centering
	\includegraphics[width=0.45\textwidth]{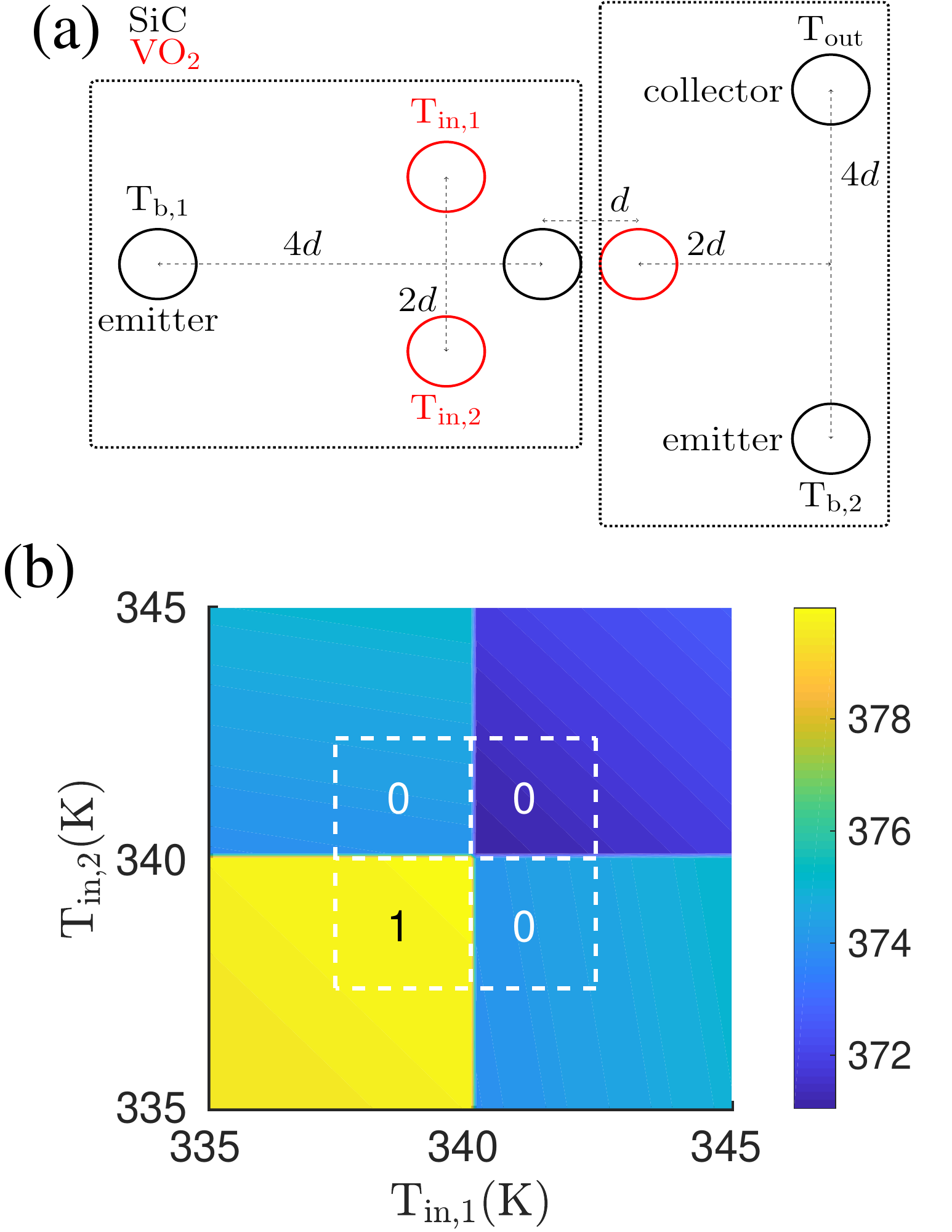}
	\caption{ {NOT + OR = NOR gate.} (a) Sketch of an OR gate coupled to a NOT gate with $T_{b,1} = 300\,{\rm K}$ and $T_{b,2} = 400\,{\rm K}$. (b) The output temperature of the NOT + OR gate has the functionality of a NOR gate. Note that this output temperature scheme can also be interpreted as a NAND gate with a slight redefinition of the range of output temperatures for ``0" and ``1".}
  \label{Fig:NOTORgate}
\end{figure}

After identifying possible configurations of nanoparticles which define fundamental logic gates, it is natural to address the problem of the combination of individual logic gates. With this respect, the issue of non-additivity of radiative heat flux is expected to play a role. We quantitatively address this problem in this section. To this aim, we start from the simple combination of an OR gate and a NOT gate as sketched in Fig.~\ref{Fig:NOTORgate}(a). The emitter and collector of the NOT gate have this time a relatively large distance of 400\,nm whereas the base of the NOT gate is shifted towards the collector of the OR gate such that the distance between them is 100\,nm. In this way, we tune the relative coupling strengths such that the effect is relative large. The output temperature of the combined gates, shown in Fig.~\ref{Fig:NOTORgate}(b), has indeed the signature of a NOR, with a threshold temperature for the output temperature around 375\,K. But we have to emphasize that this functionality is due to the fact that the heat flux coming from $T_{b,2}$ is partially absorbed from the input terminals and the bath $T_{b,1}$. When $T_{\text{in},1} > T_c$ or $T_{\text{in},2} > T_c$, this effect increases, cooling down the output particle. In particular, the base particle of the NOT gate is not undergoing any phase transition but stays in its metallic phase. When making the distance between the emitter of the NOT gate and the base larger, then we can bring the base particle to a temperature regime where it makes a phase transition. But in this case, we find that the NOR gate functionality will be lost. This is due to the fact that making the phase transition does not only change the coupling within the NOT gate but also changes the coupling to the particles of the OR gate and therefore results in a completely different configuration. We further want to emphasize that the result in Fig.~\ref{Fig:NOTORgate}(b) can also be interpreted as a NAND gate when redefining the range of output temperatures for ``0" and ``1" slightly. Therefore there is some ambiguity. We have tried several other ways of combining individual gates, but it seems that a naive parallel or linear connection does not necessarily result in the expected output. Hence, the main message here is that a circuitry with the basic elements of thermotronics is not obvious. The non-additive nature of radiative heat transfer makes it necessary to design a specific combination of heat-exchanging bodies for each logic output we need to realize.

\begin{figure}
	\centering
	\includegraphics[width=0.4\textwidth]{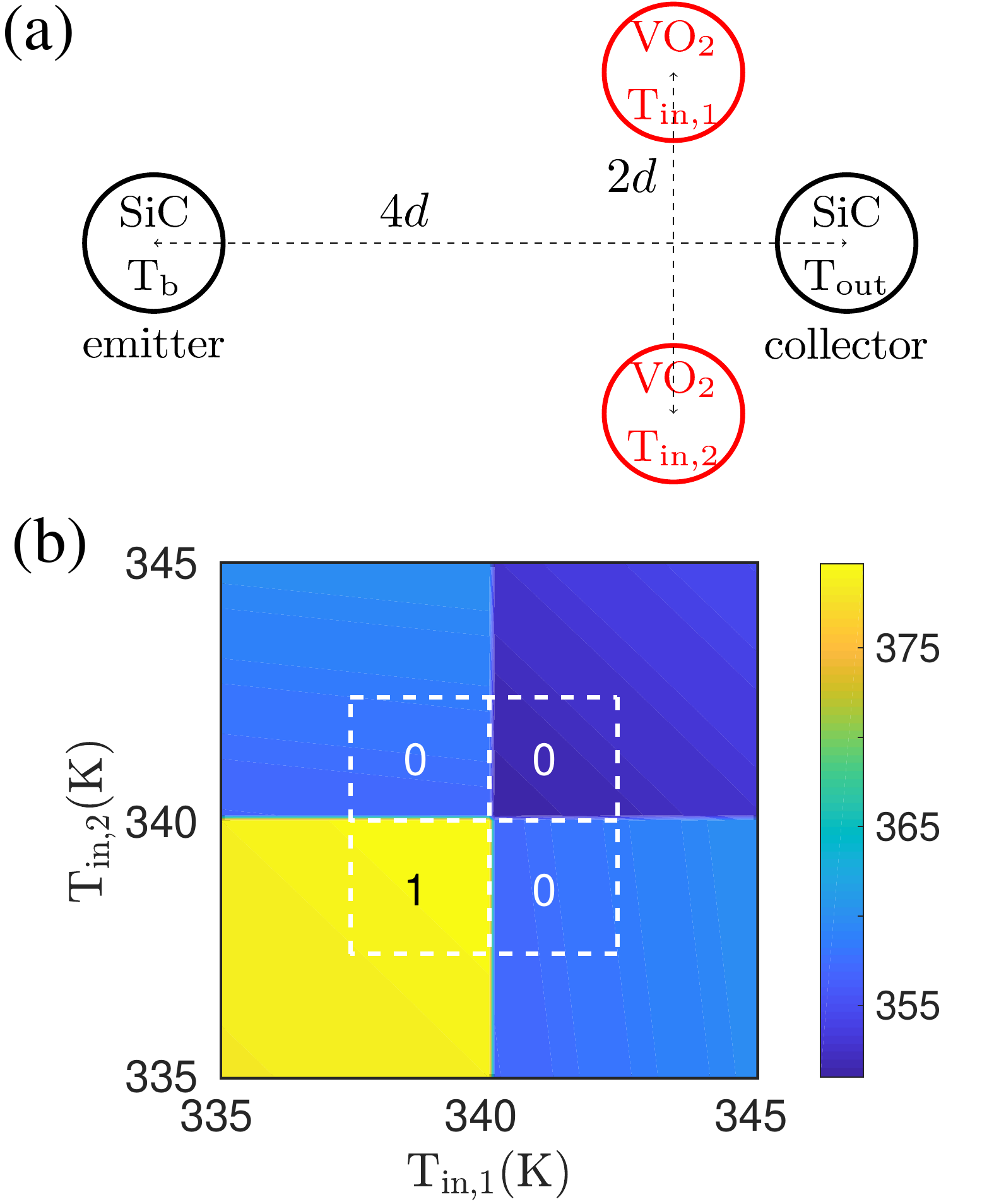}
	\caption{{NOR gate.} (a) Sketch of the NOR gate configuration. The radii of the nanoparticles is $a = 25\,{\rm nm}$ and the interparticle distance between the two input particles is $2d = 200\,{\rm nm}$ and between the emitter and collector particle the distance is $4d = 400\,{\rm nm}$ and $T_b = 400\,{\rm K}$. (b) Output temperature as function of the two input temperatures $T_{\rm in, 1/2}$.}
  \label{Fig:NORgate}
\end{figure}

In the specific case of a NOR gate, we can realize it by using the same idea allowing us to design an OR gate starting from a UNIT one. We can simply use two NOT gates in parallel, by replacing its input particle by two input particles as shown in Fig.~\ref{Fig:NORgate}(a). In this configuration the temperature of the output particle is shown in Fig.~\ref{Fig:NORgate}(b) as a function of the two input temperatures. In analogy with what we have seen before for the UNIT/OR gates, the functionalities of the NOT gate works for the two input particles separately, the output is clearly that of a NOR gate if we define the $T_{\rm out} > 370\,{\rm K}$ as ``1" and  $T_{\rm out} < 370\,{\rm K}$ as ``0". It should be noted that by a slightly different definition of the output temperature deciding between ``0" and ``1" the gate in Fig.~\ref{Fig:NORgate}(b) could as well be interpreted as a NAND gate.

We conclude by giving one clear example of failure of the simple combination of logic gates due to non additivity of radiative heat transfer. As a matter of fact, if we realize a combination of gates as in Fig.~\ref{Fig:NOTORgate}(a), but with $T_{b,1} = 400\,{\rm K}$ and $T_{b,2} = 400\,{\rm K}$, i.e. we combine a NOR and a NOT gate, we obtain again the temperature scheme of a NOR gate (not shown), and not the OR we would expect. This confirms once more the need for an appropriate design of the desired logical operation and the limitations induced by non additivity.

\subsection{AND and NAND gate}

\begin{figure}
	\centering
	\includegraphics[width=0.4\textwidth]{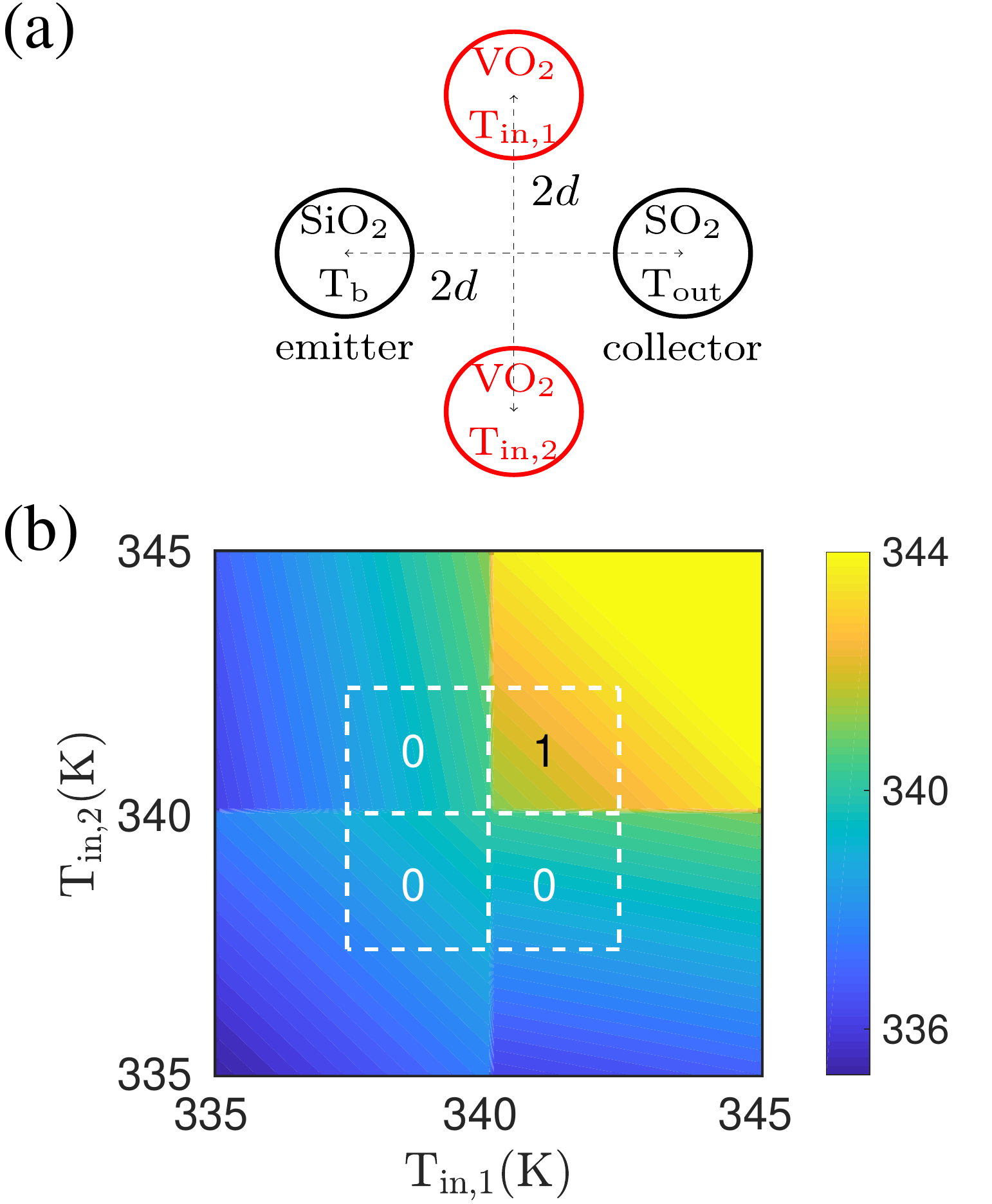}
	\caption{{AND gate.} (a) Sketch of the AND gate configuration. The radii of the nanoparticles is $a = 25\,{\rm nm}$ and the interparticle distance between the two input particles and between the emitter and collector particle is $2d = 200\,{\rm nm}$ and $T_b = 400\,{\rm K}$. (b) Output temperature as function of the two input temperatures $T_{\rm in, 1/2}$.}
  \label{Fig:ANDgate}
\end{figure}

\begin{figure}
	\centering
	\includegraphics[width=0.6\textwidth]{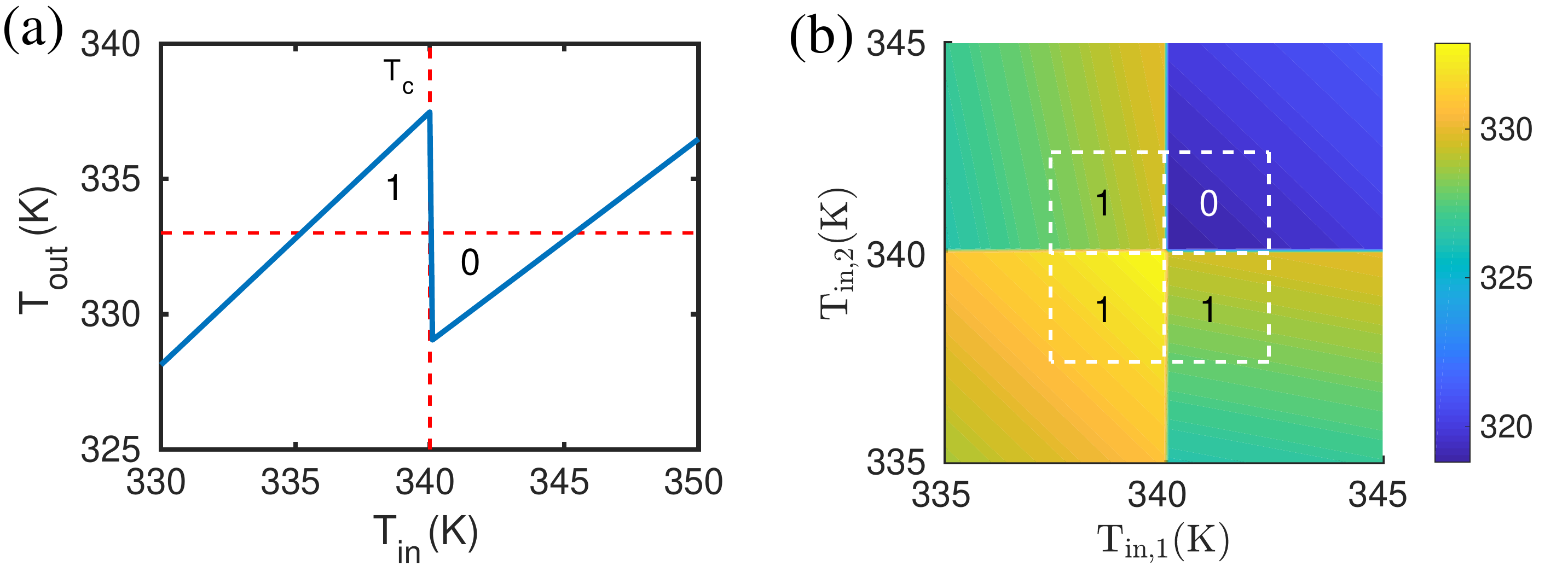}
	\caption{{NOT and NAND gate.} (a) The output temperature of the configuration in Fig.~\ref{Fig:NOTgateGlass}(a) with $T_b = 300\,{\rm K}$. The functionality is that of a NOT gate for logical operations. (b) The output temperature of the configuration in Fig.~\ref{Fig:ANDgate}(a)  with $T_b = 300\,{\rm K}$. The functionality is that of a NAND gate.}
  \label{Fig:NANDgate}
\end{figure}

As can be expected from the previous discussion of the NOR gate, we can expect to realize an AND gate by putting two UNIT gates in parallel, which amounts to replace the SiC particles in the NOR gate by SiO$_2$ particles. Indeed, as shown in Fig.~\ref{Fig:ANDgate} this results in an AND gate. In the same manner as before we can invert the functionality of the UNIT gate and therefore also of the AND gate by using $T_b = 300\,{\rm K}$ instead of $T_b = 400\,{\rm K}$. The resulting NOT gate using SiO$_2$ particles and the NAND gate are shown in Fig.~\ref{Fig:NANDgate}.

\subsection{Impact of the width of the VO$_2$ phase-transition region}

In the previous calculations we neglected that the phase transition extends over a range of several degrees and we have further neglected the hysteresis~\cite{OrdonezEtAl2016}. The impact of the extended region for the phase transition will of course limit the region where we can clearly define the ``0" or ``1" to the region outside the phase transition region. Furthemore, due to the hysteresis effect, this transition region will be slightly shifted when cooling the VO$_2$ particles compared to the case where they are heated. Therefore, depending on the way the logic gates are used, the temperature range in which the gate operates will be slightly different. A detailed analysis of the dynamics of the logic gates, the impact of the extended transition region and the hysteresis effect is out of the scope of the present work and will be done elsewhere.

\begin{figure}
	\centering
	\includegraphics[width=0.6\textwidth]{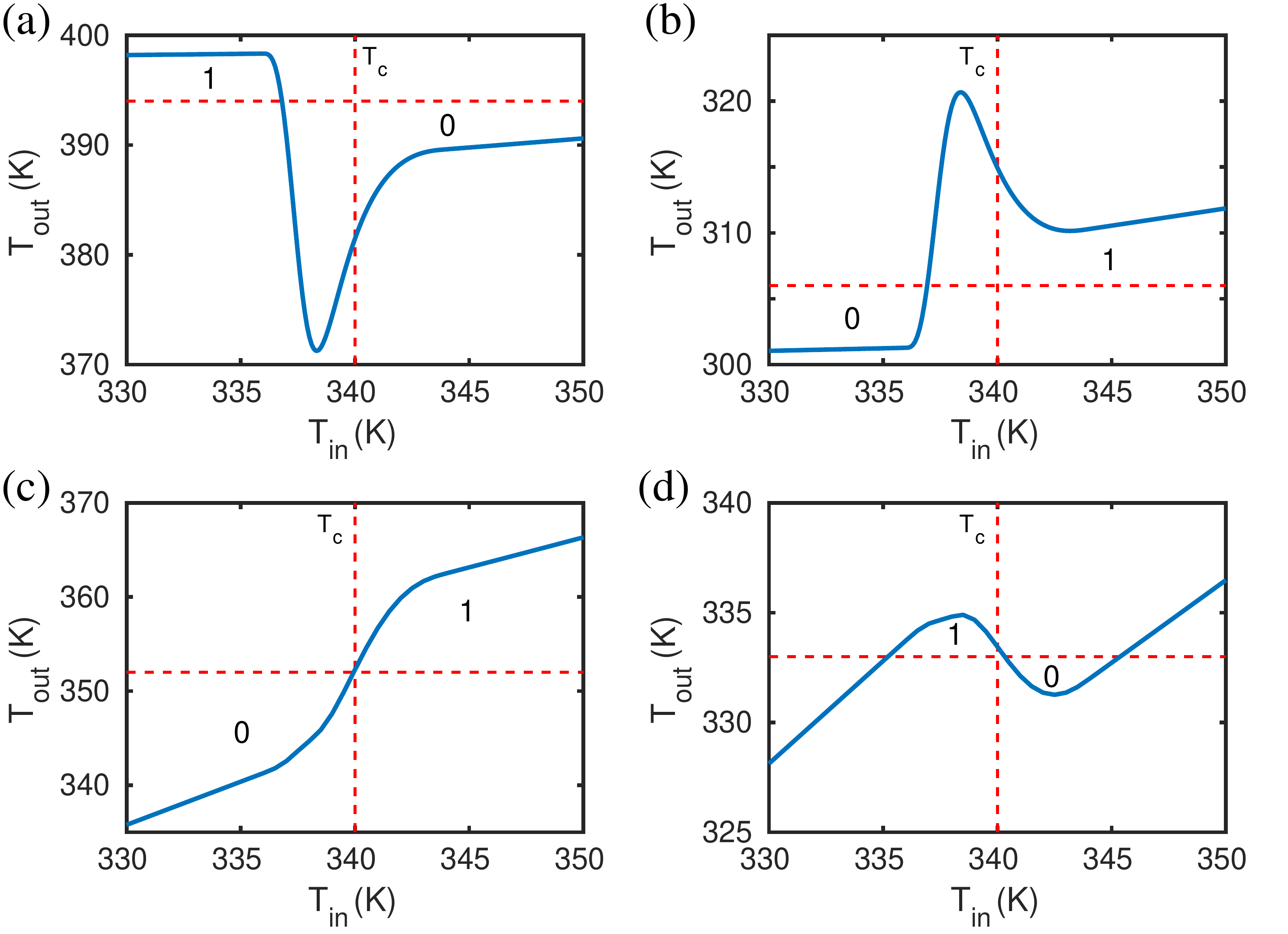}
	\caption{{Finite phase transition region.} NOT and UNIT gates with a finite transition region for $T \in [336\,{\rm K}, 344\,{\rm K}]$. The gates with SiC [see (a) and (b)]  as output particles are the same as configurations corresponding to Figs.~\ref{Fig:NOTgate}(b) and \ref{Fig:Fig4}(b) and the gates with SiO$_2$ [see (c) and (d)] as output particles are the same configurations corresponding to Figs.~\ref{Fig:NANDgate}(a) and \ref{Fig:NOTgateGlass}(b).}
  \label{Fig:smoothgates}
\end{figure}

Here for convenience we illustrate the effect of an extended phase transition region only for the NOT and UNIT gates realized with SiC and SiO$_2$, respectively. To this end, we model the permittivity of VO$_2$ in the transition region by the effective medium expression
\begin{equation}
  \epsilon_{{\rm VO}_2}^{\rm eff} = (1 - f) \epsilon_{{\rm VO}_2}^{\rm d}  + f \epsilon_{{\rm VO}_2}^{\rm m},
\end{equation}
where $\epsilon_{{\rm VO}_2}^{\rm d/m}$ are the permittivities of VO$_2$ in the dielectric/metallic phase. The filling factor $f \in [0,1]$ can be modeled in different ways. Here we use a so called smoother step function
\begin{equation}
 f(x) = 6x^5-15x^4+10x^3
\end{equation}
with $x \in [0,1]$ realising a smooth transition from 0 to 1. To model the transition in the temperature region $T \in [336\,{\rm K}, 344\,{\rm K}]$ we use $x = (T - 336\,{\rm K})/8\,{\rm K}$. The results for the NOT and UNIT gates are shown in Fig.~\ref{Fig:smoothgates}. 

For the SiO$_2$-based gates in  Fig.~\ref{Fig:smoothgates}(c) and  Fig.~\ref{Fig:smoothgates}(d) we simply obtain a smoothed transition of the jumps at $T_C$ observed in Fig.~\ref{Fig:NOTgateGlass} and Fig.~\ref{Fig:NANDgate}(a). Interestingly, for the SiC-based gates we have a non-monotonous transition in Fig.~\ref{Fig:smoothgates}(a) and Fig.~\ref{Fig:smoothgates}(b). It turns out that there is an optimal coupling at a filling fraction of $f = 0.18$ which corresponds in our model for $f(x)$ to a temperature of $338.5\,{\rm K}$. Due to this optimal coupling there is a minimum in the NOT gate in  Fig.~\ref{Fig:smoothgates}(a) and a maximum in the UNIT gate in Fig.~\ref{Fig:smoothgates}(b) because at this filling fraction the input VO$_2$ particle optimally cools or heats the SiC output particle, resp. Anyway, when defining the transition temperature defining ``0'' and ``1'' of the output by the vertical dashed lines  in Fig.~\ref{Fig:smoothgates}(a) and Fig.~\ref{Fig:smoothgates}(b) we find that the modelling of the finite transition region shifts the temperature at which the gates switch to temperatures smaller than $T_c$. In our model the switching of the gates happens at approximately $338\,{\rm K}$ instead of $T_c = 340\,{\rm K}$.

\section{Conclusions}

We have introduced here basic scalable circuits to make NOT, OR, NOR, AND and NAND gates from heat flux exchanged in near-field regime in simple nanoparticle networks made with phase change materials. Like in electronics, the NOR gate is obtained by a parallel arrangement of NOT gates. We have shown that the NOT, NOR and AND gate functionalities can be inverted by simply tuning the temperature of surrounding bath around the critical temperature of phase change material.

The development of thermal circuits based on the combination of logic gates should open the door to innovative solutions for an active thermal management of heat flux generated at nanoscale in solid materials. In a long-term perspective, this thermotronics should allow the development of smart systems to take decisions and act using heat flux rather than electricity and even to allow machine-to-machine communication with heat. Potential applications are the zero-electricity presence and movement monitoring of an object using thermal sensors and actuators exposed to the infrared radiation coming from the object. This signal could sequentially activate logic gates, launching a series of boolean operations to analyze the thermal signals, without any other external power making this technology autonomous.  For information processing this speed is obviously not competitive with the current electronics devices but it is more than enough for active thermal management and thermal sensing. Moreover, the use of 2D materials with extremely weak heat capacity or systems out of thermal equilibrium could in the next few years drastically reduce the operating speed of thermal circuits.


\subsection*{Author contributions}

C.~K. performed the numerical simulations. All the authors discussed the results and contributed to the preparation of the manuscript.

\subsection*{Acknowledgments}

S.-A.\ B. acknowledges support from Heisenberg Programme of the Deutsche Forschungsgemeinschaft (DFG, German Research Foundation) under the project No. 404073166.

\subsection*{Competing interests}

The authors declare no competing interests.

\end{document}